\pdfoutput=1

\PassOptionsToPackage{inline}{enumitem}
\PassOptionsToPackage{dvipsnames}{xcolor}

\RequirePackage{expl3}
\documentclass[futureinternet,article,accept,moreauthors,pdftex,10pt,a4paper]{Definitions/mdpi}

\usepackage{iftex}

\ifPDFTeX%
  \usepackage[utf8]{inputenc}
\else
  \RequirePackage{fontspec}
\fi

\usepackage{amsmath,amssymb,amsfonts}
\usepackage{algorithmic}
\usepackage{textcomp}

\usepackage{caption}
\usepackage{csquotes}
\usepackage{enumitem}
\usepackage{etoolbox}
\usepackage{layouts}
\usepackage{subcaption}
\usepackage{xcolor}

\usepackage[labelformat=simple]{subcaption}

\DeclareCaptionLabelFormat{subcaptionlabel}{\normalfont(\textbf{#2}\normalfont)}
\graphicspath{
    {images/}
    {template/MDPI/}
}

\firstpage{1}
\pubvolume{xx}
\issuenum{1}
\articlenumber{1}
\pubyear{2018}
\copyrightyear{2018}
\history{Received: 6 May 2018; Accepted: 12 June 2018; Published: 15 June 2018}

\newcommand{\abs}[1]{ \left\lvert#1\right\rvert}

\newcommand{\adamIncludeFigure}[3]{
  \subcaptionbox{#2}{\includegraphics[width=#1\linewidth]{#3}}
}

\newcommand{\adamIncludeFigureCS}[4]{
  \subcaptionbox{#3}[#2\linewidth]{\includegraphics[width=#1\linewidth]{#4}}
}

\Title{StegNet: Mega Image Steganography Capacity with Deep Convolutional Network}

\Author{%
  Pin Wu  \(^{1}\)  \orcidA, Yang Yang \(^{1}\) \orcidB\,and Xiaoqiang Li \(^{1,2,}\)* \orcidC}

\AuthorNames{%
    Pin Wu
,   Yang Yang
and Xiaoqiang Li
}

\address{%
\(^{1}\) \quad School of Computer Science, Shanghai University, China; wupin@shu.edu.cn (P.W.); adamcavendish@shu.edu.cn (Y.Y.)\\
\(^{2}\) \quad Shanghai Institute for Advanced Communication \& Data Science, Shanghai University, China
}

\corres{Correspondence: xqli@shu.edu.cn}

\abstract{Traditional image steganography often leans interests towards safely embedding hidden information into cover images with payload capacity almost neglected. This paper combines recent deep convolutional neural network methods with image-into-image steganography. It successfully hides the same size images with a decoding rate of 98.2\% or bpp (bits per pixel) of 23.57 by changing only 0.76\% of the cover image on average. Our method directly learns end-to-end mappings between the cover image and the embedded image and between the hidden image and the decoded image. We~further show that our embedded image, while with mega payload capacity, is still robust to statistical analysis.
}

\begin{document}

\section{Introduction}%
\label{sec:introduction}

Image steganography, aiming at delivering a modified cover image to secretly transfer hidden information inside with little awareness of the third-party supervision, is a classical computer vision and cryptography problem. Traditional image steganography algorithms go to their great length to hide information into the cover image while little consideration is tilted to payload capacity, also known as the ratio between hidden and total information transferred. The~payload capacity is one significant factor to steganography methods because if more information is to be hidden in the cover, the~visual appearance of the cover is altered further and thus the risk of detection is higher 
(The source code is available at: \url{https://github.com/adamcavendish/StegNet-Mega-Image-Steganography-Capacity-with-Deep-Convolutional-Network}).

The most commonly used image steganography for hiding large files during transmission is embedding a RAR archive (Roshal ARchive file format) after a JPEG (Joint Photographic Experts Group) file. In such way, it can store an infinite amount of extra information theoretically. However, the~carrier file must be transmitted as it is, since any third-party alteration to the carrier is going to destroy all the hidden information in it, even just simply read out the image and save it again will corrupt the hidden information.

\begin{figure}[H]
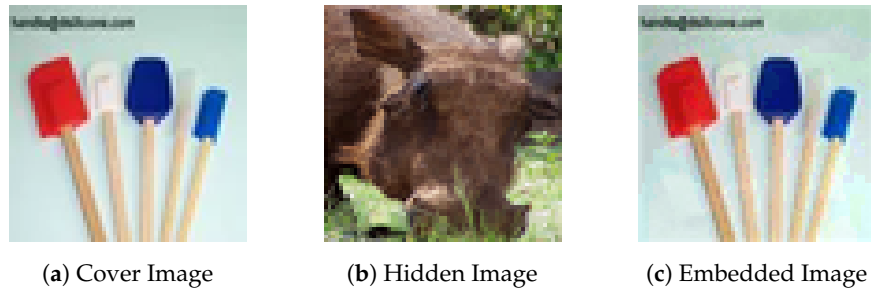

  \centering
  \begin{tabular}{ccc}
    \adamIncludeFigureCS{0.2}{0.25}{Cover Image}{vague_outline/hidden_outline_visible_covr.png}
    \adamIncludeFigureCS{0.2}{0.25}{Hidden Image}{vague_outline/hidden_outline_visible_hide.png}
    \adamIncludeFigureCS{0.2}{0.25}{Embedded Image}{vague_outline/hidden_outline_visible_steg_lsb4.png}
  \end{tabular}
  \vspace{-8pt}
  \caption{Vague Outline Visible in 4-bit LSB Steganography Embedded-Cover-Diversity = 50\%, Hidden-Decoded-Diversity = 50\%, Payload Capacity = 12 bpp.}%
\label{fig:vagueoutline}
\end{figure}

To maximize the payload capacity while still resistible to simple alterations, pixel level steganography is majorly used, in which LSB (least significant bits) method~\cite{LSBRevisited}, BPCS~\cite{BPCS} (Bit Plane Complexity Segmentation), and their extensions are in dominant. LSB-based methods can achieve a payload capacity of up to 50\%, or otherwise, a vague outline of the hidden image would be exposed (see Figure~\ref{fig:vagueoutline}). However, most of these methods are vulnerable to statistical analysis, and therefore it can be easily detected.

Some traditional steganography methods with balanced attributes are hiding information in the JPEG DCT components. For instance, A. Almohammad's work~\cite{HCJPEG} provides around 20\% of payload capacity (based on the patterns) and still remains undetected through statistical analysis.

Most secure traditional image steganography methods recently have adopted several functions to evaluate the embedding localizations in the image, which~enables content-adaptive steganography. HuGO~\cite{HuGO} defines a distortion function domain by giving every pixel a changing cost or embedding impact based on its effect. It uses a weighted norm to represent the feature space. WOW (Wavelet Obtained Weights)~\cite{WOW} embeds information according to the textural complexity of the image regions. Work~\cite{UniDistortion, CASMinStatDetect} have discussed some general ways of content-adaptive steganography to avoid statistical analysis. Work~\cite{CASBatch} is focusing on content-adaptive batched steganography. These methods highly depend on the patterns of the cover image, and therefore the average payload capacity can be hard to~calculate.

\begin{figure}[H]
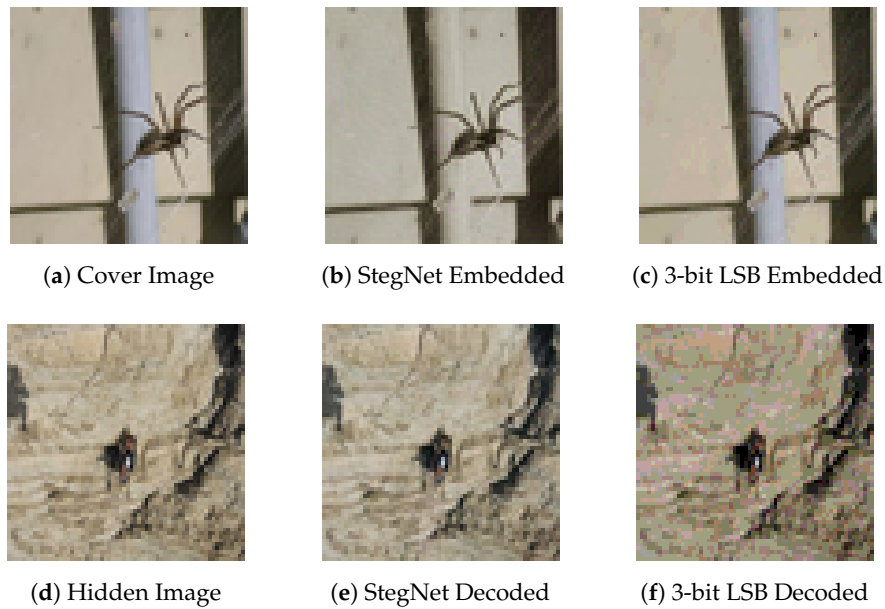

  \centering
  \begin{tabular}{ccc}
    \adamIncludeFigureCS{0.2}{0.25}{Cover Image}        {effect_comparison/image_1_covr.png}
    \adamIncludeFigureCS{0.2}{0.25}{StegNet Embedded}   {effect_comparison/image_1_steg_stegnet.png}
    \adamIncludeFigureCS{0.2}{0.25}{3-bit LSB Embedded} {effect_comparison/image_1_steg_lsb3.png}\\

    \adamIncludeFigureCS{0.2}{0.25}{Hidden Image}       {effect_comparison/image_1_hide.png}
    \adamIncludeFigureCS{0.2}{0.25}{StegNet Decoded}    {effect_comparison/image_1_dcpt_stegnet.png}
    \adamIncludeFigureCS{0.2}{0.25}{3-bit LSB Decoded}  {effect_comparison/image_1_dcpt_lsb3.png}
  \end{tabular}
  \vspace{-8pt}
  \caption{{~StegNet and 3-bit LSB Comparison Embedded-Cover-Diversity = 0.76\%, Hidden-Decoded- Diversity = 1.8\%, Payload Capacity = 23.57 bpp.}}%
\label{fig:stegnetvslsb3}
\end{figure}

The major contributions of our work are as follows:
\begin{enumerate*}[label=\roman*)]
  \item[(i)] We propose a methodology to apply neural networks for image steganography to embed image information into image information without any help of traditional steganography methods.
  \item[(ii)] Our implementation raises image steganography payload capacity to an average of 98.2\% or 23.57 bpp (bits per pixel), changing only around 0.76\% of the cover image (See Figure~\ref{fig:stegnetvslsb3}).
  \item[(iii)] We propose a new cost function named variance loss to suppress noise pixels generated by generator network.
  \item[(iv)] Our implementation is robust to statistical analysis and 4 other widely used steganography analysis methods.
\end{enumerate*}

The decoded rate is calculated by
\csdef{CE}{\mathrm{CE}}
\csdef{HD}{\mathrm{HD}}
\begin{equation}
\textrm{Decoded Rate} = 1 - \frac{\sum_{i=1}^{N} \sum_{j=1}^{M} \abs{H_{i,j} - D_{i,j}}}{N \times M} ,
\end{equation}
the cover changing rate is calculated by
\begin{equation}
\textrm{Cover Changing Rate} = \frac{\sum_{i=1}^{N} \sum_{j=1}^{M} \abs{C_{i,j} - E_{i,j}}}{N \times M}
\end{equation}
and the bpp (bits per pixel) is calculated by
\begin{equation}
\textrm{Capacity} = \textrm{Decoded Rate} \times 8 \times 3 \quad \textrm{(bpp)}
\end{equation}
where \(C, H, E, D\) symbols stand for the cover image (\(C\)), the~hidden image (\(H\)), the~embedded image~(\(E\)) and the decoded image (\(D\)) in correspondence, and ``8, 3'' stands for number of bits per channel and number of channels per pixel respectively.

\csundef{HD}
\csundef{CE}

This paper is organized as follows. Section~\ref{sec:relatedwork} will describe traditional high-capacity steganography methods and the convolution neural network used by this paper. Section~\ref{sec:convsteg} will unveil the secret why the neural network can achieve the amount of capacity encoding and decoding images. The~architecture and experiments of our neural network are discussed in Sections~\ref{sec:architecture} and \ref{sec:experiments}, and finally, we'll make a conclusion and put forward some future works in Section~\ref{sec:conclusion}.

\section{Related Work}%
\label{sec:relatedwork}

\vspace{-6pt}
\subsection{Steganography Methods}%
\label{ssec:stegmethods}

Most steganography methods can be grouped into three basic types, which~is image domain steganography, transform domain steganography and file-format-based steganography. Image domain ones have an advantage of simplicity and better payload capacity while being more likely to be detected. Transform domain ones usually have a more complex algorithm but hides pretty well through third-party analysis. File-format-based ones depend very much on the file format which makes it quite fragile to alterations.

\subsection{JPEG RAR Steganography}%
\label{ssec:jpegrar}

The JPEG RAR Steganography is a kind of file-format-based steganography, which~uses a feature in these two file format specifications. (JPEG~\cite{jpegspec} and RAR~\cite{rarspec})

After the JPEG file has scanned the segment of EOI (End Of Image) (0xd9 in hex format), all the remaining segments are ignored (skipped), and therefore any information is allowed to be appended afterward. A RAR file~~\cite{rarspec} has the magic file header ``0x52 0x61 0x72 0x21 0x1a 0x07 0x00'' in hex format (\textquote{Rar!} as characters) and the parser will ignore all the information before the file header. It is possible to dump the binary of the RAR file after the JPEG file, and it'll apparently act as if it is a JPEG image file while it is actually also a RAR archive. However, the~method is very fragile to any file alterations. Third-party surveillance might truncate useless information to save transmission resource or apply some image alterations to attack potential steganography. Any alteration will crash the steganography, and all hidden information is lost.

\subsection{LSB (Least Significant Bit) Method}%
\label{ssec:lsbmethod}

\begin{figure}[H]
  \centering
  \includegraphics[width=0.5\linewidth]{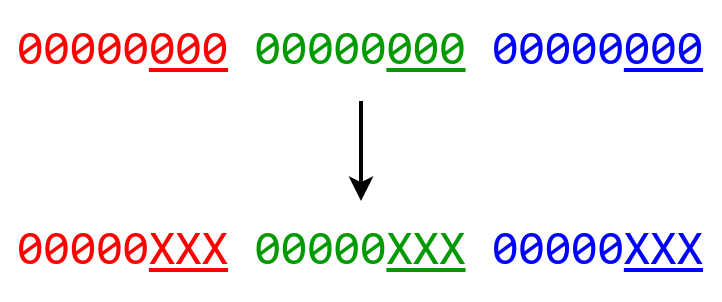}
  \caption{LSB Explaination.}%
\label{fig:lsbexplained}
\end{figure}

LSB (Least Significant Bit)-based methods~\cite{LSBRevisited} are the most commonly used image domain steganography methods which hide information at the pixel level. Most LSB methods aim at altering parts of the cover image to such an extent that human visual system can barely notice. These methods are motivated by the fact that the visual part of most figures is dominated by the highest bits of each pixel, and the LSB bits (the underlined part of one pixel as shown in Figure~\ref{fig:lsbexplained}) are statistically similar to randomly generated data, and therefore, hiding information via altering LSB cannot change the visual result apparently.

The embedding operation of LSB method for the least bit single channel image is described as~follows:
\begin{equation*}
S_{i} = (C_{i} \land \mathrm{FE_{HEX}}) \lor ((M_{i} \land \mathrm{80_{HEX}}) >> 7)
\end{equation*}
where \(S_{i}\), \(C_{i}\) and \(M_{i}\) are the \emph{i}th pixel of image after steganography, \emph{i}th pixel of cover image and \emph{i}th bit of the hiding message.

Since the least significant bits of the image data should look like random data, there are major two schemes in distributing the hiding data. The~first kind of methods is to put in the hiding message sequentially after encrypting or compressing to achieve the randomness. The~second kind of methods is scattering the hiding data by adopting a mutually acknowledged random seed by which generates the actual hiding sequence~\cite{RobustImageSteg}.

\subsection{JPEG Steganography}%
\label{ssec:hcjpeg}

JPEG steganography, i.e., Chang's work~\cite{JPEGSteg} and A. Almohammad's work~\cite{HCJPEG}, is a part of transform domain steganography. JPEG format examines an image in \(8 \times 8\) blocks of pixels, converts from RGB color space into YCrCb (luminance and chrominance) color space, applies DCT (Discrete Cosine Transformation), quantizes the result, and entropy encodes the rest. After lossy compression, which~is after quantization, the~hidden information is hidden into the quantized DCT components, which~serves as an LSB embedding in the transformed domain. As a result, it is quite hard to detect using statistical analysis and comparably lower payload capacity to LSB method.

\subsection{Convolutional Neural Network}%
\label{ssec:convnet}

Convolutional neural network~\cite{conv}, though dates back to the 1990s, is now trending these years after AlexNet~\cite{alexnet} won the championship of ImageNet competition. It has successfully established new records in many fields like classification~\cite{imagenet2017}, object segmentation~\cite{coco2016}, etc. A lot of factors boosted the progress including the development of modern GPU hardware, the~work of ReLU (Rectified Linear Unit)~\cite{relu} and its extensions, and finally, the~abundance of training data~\cite{imagenet}. Our work also benefits a lot from these factors.

The convolution operation is not solely used in neural networks, but also widely used in traditional computer vision methods. For instance, gaussian smoothing kernel is extensively used for image blurring and noise reduction, which, in implementation, is equal to applying a convolution between the original image and a gaussian function. Many other contributions in traditional methods are handcrafted patterns, kernels or filter combinations, i.e.,\ the Sobel-Feldman filter~\cite{SobelFeldmanFilter} for edge detection, Log-Gabol filter~\cite{fischer07cv} for texture detection, HOG~\cite{HOG} for object detection, etc.

However, designing and tuning handcrafted patterns are highly technical and might be effective for only some tasks. On the contrary, convolutional neural networks have the advantage of automatically creating patterns for specific tasks through back-propagation~\cite{RumelhartBP} on its own, and even further, high-level features can be easily learned through combinations of convolution operations~\cite{Zeiler_Fergus_2013, olah2017feature,Mahendran_Vedaldi_2014}.

\subsection{Autoencoder Neural Network}%
\label{ssec:autoencoder}

Our method is inspired by traditional autoencoder neural networks~\cite{hinton1994autoencoders}, which~was originally trained to generate an output image the same as input image in appearance. It is usually made up of two neural networks, one encoding network \(h = f(x)\) and one decoding network \(d = g(h)\), restricted under \(d = x\), who finally can learn the conditional probability distribution of \(p(h|x)\) and \(p(x|h)\) correspondently. The~autoencoder architecture has shown the ability to extract salient features in from images seen through shrinking hidden layer (\(h\))'s dimension, which~has been applied to various fields, i.e.,\ denoising~\cite{vincent2008extracting}, dimension reduction~\cite{wang2014generalized}, image generation~\cite{VAE}, etc.

\subsection{Neural Network for Steganography}%
\label{ssec:nnsteg}

Recently there are some works on applying neural networks for steganography. El-Emam~\cite{El-emam_2008} and Saleema~\cite{Saleema_Amarunnishad_2016} work on using neural networks to refine the embedded image generated via traditional steganography methods, i.e., LSB method. Volkhonskiy's~\cite{SGAN} and Shi's~\cite{SSGAN} work focus on generating secure cover images for traditional steganography methods to apply image steganography. Baluja~\cite{Baluja_2017} is working on the same field as StegNet. However, the~hidden image is slightly visible on residual images of the generated embedded images. Moreover, his architecture uses three networks which requires much more GPU memory and takes more time to embed.

\section{Convolutional Neural Network for Image Steganography}%
\label{sec:convsteg}

\vspace{-6pt}
\subsection{High-order Transformation}%
\label{ssec:highordertrans}

In image steganography, we argue that we should not only focus on where to hide information, which~most traditional methods work on, but we should also focus on how to hide it.

Most traditional steganography methods usually directly embed hidden information into parts of pixels or transformed correspondances. The~transformation regularly occurs in \textit{where to hide}, either~actively applied in the steganography method or passively applied because of file format. As a result, the~payload capacity is highly related and restricted to the area of the texture-rich part of the image detected by the \textit{handcoded} patterns.

DCT-based steganography is one of the most famous transform domain steganography. We~can consider the DCT process in JPEG lossy compression process as a kind of one-level high-order transformation which works at a block level, converting each \(8 \times 8\) or \(16 \times 16\) block of pixel information into its corresponding frequency-domain representation. Even hiding in DCT transformed frequency-domain data, traditional works~\cite{JPEGSteg, HCJPEG} embed hidden information in mid-frequency band via LSB-alike methods, which~eventually cannot be eluded.

\begin{figure}[H]
  \centering
  \includegraphics[width=0.2\linewidth]{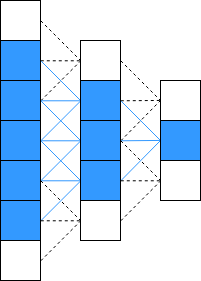}
  \caption{Receptive Field of Convolutional Neural Network.}%
\label{fig:ConvReceptiveField}
\end{figure}

While in contrast, deep convolution neural network makes multi-level high-order transformations possible for image steganography. Figure~\ref{fig:ConvReceptiveField} shows the receptive field of one high-level kernel unit in a demo of a three-layer convolutional neural network. After lower-level features are processed by kernels and propagated through activations along middle layers, the~receptive field of final higher-level kernel unit is able to absorb 5 lower-level features of the first layer and form its own higher-level feature throughout the training process.

\subsection{Trading Accuracy for Capacity}%
\label{ssec:tradingaccforcap}

Traditional image steganography algorithms mostly embed hidden information as it is or after applying lossless transformations. After decoding, the~hidden information is extracted as it is or after the corresponding detransformations are applied. Therefore, empirically speaking, it is just as file compression methods, where lossless compression algorithms usually cannot outperform lossy compression algorithms in capacity.

We need to think in a ``lossy'' way in order to embed almost equal amount of information into the cover. The~model needs to learn to compress the cover image and the hidden image into an embedding of high-level features and converts them into an image that appears as similar as the cover image, which~comes to the vital idea of trading accuracy for capacity.

Trading accuracy for capacity means that we do not limit our model in reconstructing at a pixel-level accuracy of the hidden image, but aiming at ``recreating'' a new image with most of the features in it with a panoramic view, i.e.,\ the spider in the picture, the~pipes' position relatively correct, the~outline of the mountain, etc.

In other words, the~traditional approaches work in lossless ways, which~after some preprocessing to the hidden image, the~transformed data is crammed into the holes prepared in the cover image. However, StegNet approach decoded image has no pixel-wise relationship with the hidden image at all, or strictly speaking, there is no reasonable transformation between each pair of corresponding pixels, but the decoded image as a whole can represent the original hidden image's meaning through neural network's reconstruction.

In the encoding process, the~model needs to transform from a low-level massive amount of pixel-wise information into some high-level limited sets of featurewise information with an understanding of the figure, and come up with a brand new image similar to the cover apparently but with hidden features embedded. In the decoding process, on the contrary, the~model is shown only the embedded figure, from which both cover and hidden high-level features are extracted, and the hidden image is rebuilt according to network's own comprehension.

\begin{figure}[H]
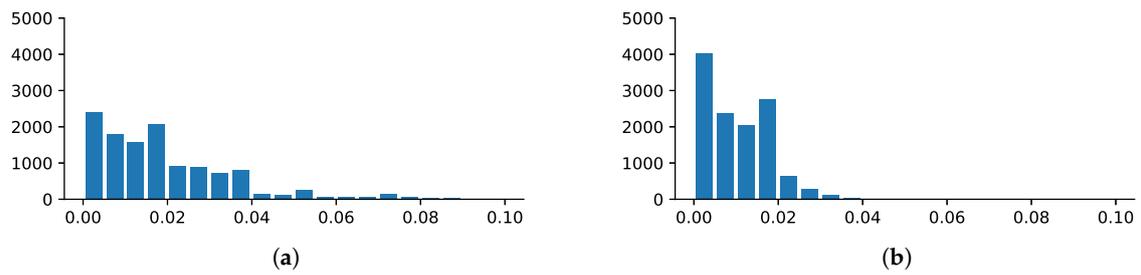

  \centering
  \begin{tabular}{cc}
    \adamIncludeFigure{0.5}{}{residual_comparison/histogram/image_1_diff_covr_steg_hist_stegnet_mag01_sp}
    \adamIncludeFigure{0.5}{}{residual_comparison/histogram/image_1_diff_hide_dcpt_hist_stegnet_mag01_sp}
  \end{tabular}
  \vspace{-16pt}
  \caption{Residual image histograms shows that the residual error is distributed  across the images. \textbf{(a)} Residual between cover and embedded; \textbf{(b)} Residual between
  hidden and decoded.}%
\label{fig:StegNetResidualHist}
\end{figure}

\begin{figure}[H]
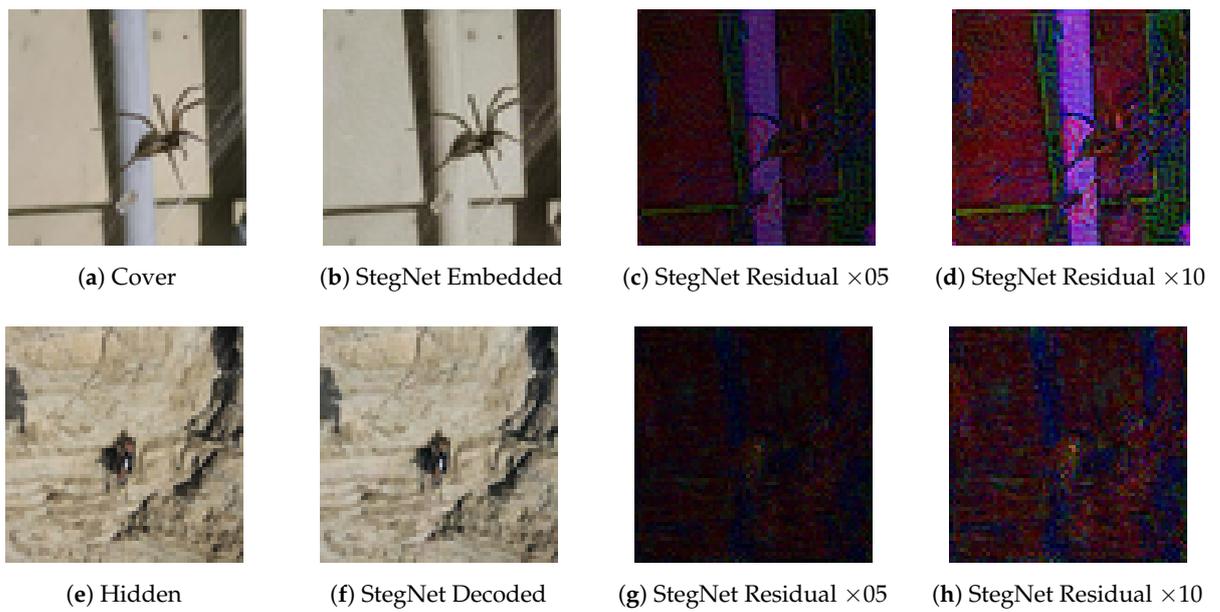

  \centering
  \begin{tabular}{cccc}
    \adamIncludeFigureCS{0.2}{0.25}{Cover}               {residual_comparison/image_1_covr.png}
    \adamIncludeFigureCS{0.2}{0.25}{StegNet Embedded}    {residual_comparison/image_1_steg_stegnet.png}
    \adamIncludeFigureCS{0.2}{0.25}{StegNet Residual \(\times 05\)}{residual_comparison/magnified/image_1_diff_covr_steg_stegnet_mag05.png}
    \adamIncludeFigureCS{0.2}{0.25}{StegNet Residual \(\times 10\)}{residual_comparison/magnified/image_1_diff_covr_steg_stegnet_mag10.png}\\

    \adamIncludeFigureCS{0.2}{0.25}{Hidden}              {residual_comparison/image_1_hide.png}
    \adamIncludeFigureCS{0.2}{0.25}{StegNet Decoded}     {residual_comparison/image_1_dcpt_stegnet.png}
    \adamIncludeFigureCS{0.2}{0.25}{StegNet Residual \(\times 05\)}{residual_comparison/magnified/image_1_diff_hide_dcpt_stegnet_mag05.png}
    \adamIncludeFigureCS{0.2}{0.25}{StegNet Residual \(\times 10\)}{residual_comparison/magnified/image_1_diff_hide_dcpt_stegnet_mag10.png}
  \end{tabular}
  \vspace{-14pt}
  \caption{StegNet residual images  ``\(\times 05\)'' and ``\(\times 10\)'' are the pixel-wise enhancement ratio.}%
\label{fig:StegNetResidualFig}
\end{figure}

\begin{figure}[H]
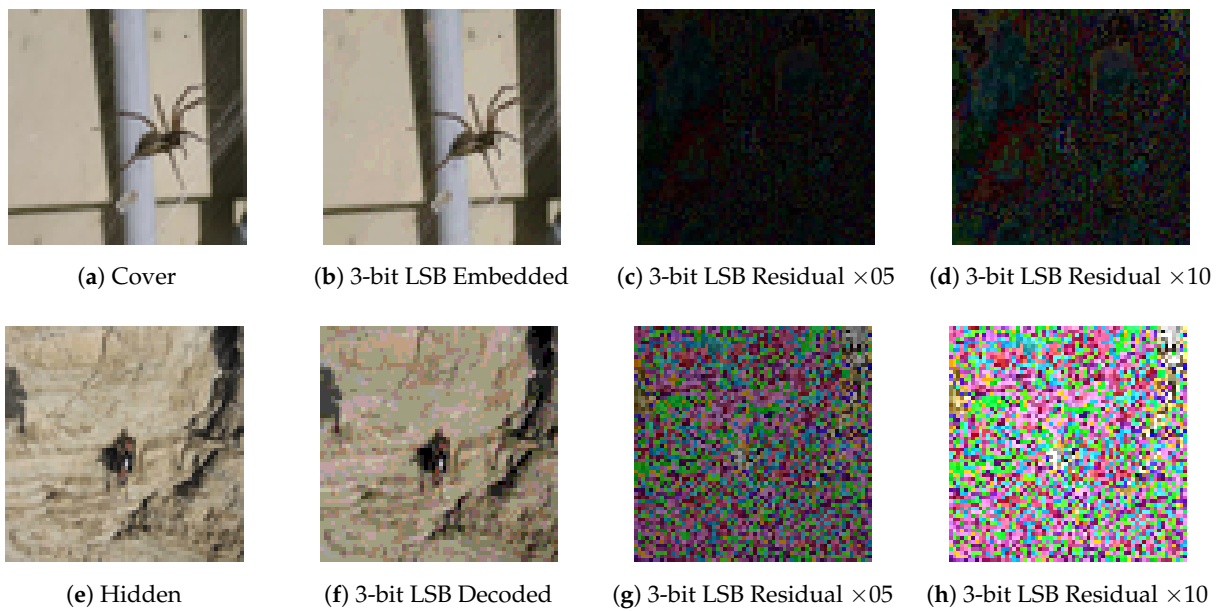

  \centering
  \begin{tabular}{cccc}
    \adamIncludeFigureCS{0.2}{0.25}{Cover}                 {residual_comparison/image_1_covr.png}
    \adamIncludeFigureCS{0.2}{0.25}{3-bit LSB Embedded}    {residual_comparison/image_1_steg_lsb3.png}
    \adamIncludeFigureCS{0.2}{0.25}{3-bit LSB Residual \(\times 05\)}{residual_comparison/magnified/image_1_diff_covr_steg_lsb3_mag05.png}
    \adamIncludeFigureCS{0.2}{0.25}{3-bit LSB Residual \(\times 10\)}{residual_comparison/magnified/image_1_diff_covr_steg_lsb3_mag10.png}\\

    \adamIncludeFigureCS{0.2}{0.25}{Hidden}                {residual_comparison/image_1_hide.png}
    \adamIncludeFigureCS{0.2}{0.25}{3-bit LSB Decoded}     {residual_comparison/image_1_dcpt_lsb3.png}
    \adamIncludeFigureCS{0.2}{0.25}{3-bit LSB Residual \(\times 05\)}{residual_comparison/magnified/image_1_diff_hide_dcpt_lsb3_mag05.png}
    \adamIncludeFigureCS{0.2}{0.25}{3-bit LSB Residual \(\times 10\)}{residual_comparison/magnified/image_1_diff_hide_dcpt_lsb3_mag10.png}
  \end{tabular}
  \vspace{-12pt}
  \caption{ 3-bit LSB residual images ``\(\times 05\)'' and ``\(\times 10\)'' are the pixel-wise enhancement ratio.}%
\label{fig:LSB3ResidualFig}
\end{figure}

As shown in Figures~\ref{fig:StegNetResidualHist} and \ref{fig:StegNetResidualFig}, StegNet is not applying LSB-like or simple merging methods to embed the hidden information into the cover. The~residual image is neither simulating random noise (LSB-based approach, see Figure~\ref{fig:LSB3ResidualFig}) nor combining recognizable hidden image inside. The~embedded pattern is distributed across the whole image and even magnified 5 to 10 times, the~residual image is similar to the cover image visually which can help decrease the abnormality exposed to the human visual system and finally avoid to be detected.

\vspace{-12pt}

\vspace{-12pt}

The residual image is computed via
\begin{equation}
  R(I_{1}, I_{2}) = \frac{\abs{I_{1} - I_{2}}}{\max \abs{I_{1} - I_{2}}} ,
\end{equation}
and the magnification or the enhancement operation is achieved via
\begin{equation}
  E(I, M) = \mathrm{clip}(I \cdot M, 0, 1) ,
\end{equation}
where \(I\) takes residual images, which~are effectively normalized to \([0, 1]\) and \(M\) is the magnification ratio, which~\(5\) and \(10\) are chosen visualize the differences in this paper.

\section{Architecture}%
\label{sec:architecture}
\vspace{-6pt}
\subsection{Architecture Pipeline}%
\label{ssec:pipeline}

\begin{figure}[H]
  \centering
  \includegraphics[width=\linewidth]{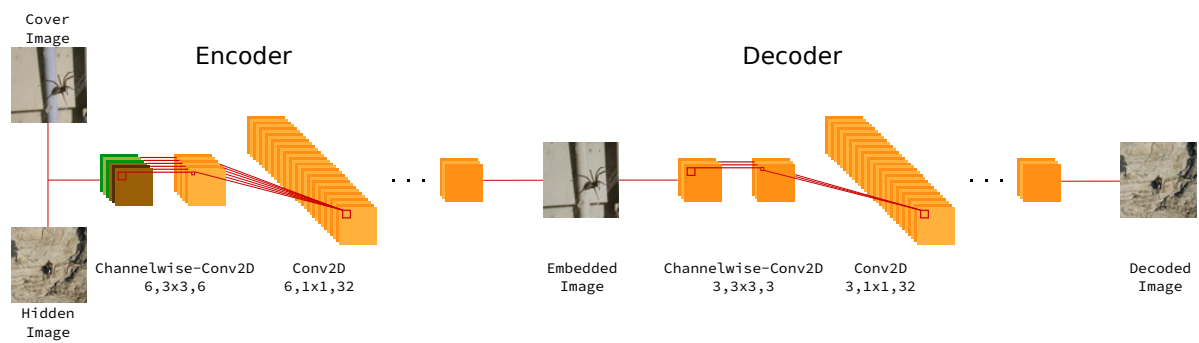}
  \vspace{-12pt}
  \caption{StegNet Processing Pipeline.}%
\label{fig:pipeline}
\end{figure}

The whole processing pipeline is shown in Figure~\ref{fig:pipeline}, which~consists of two almost identical neural network structure responsible for encoding and decoding. The~identical structures are taken from Autoencoder~\cite{hinton1994autoencoders}, GAN~\cite{GAN}, etc., which~help the neural network model similar high-level features of images in their latent space. The~details of embedding and decoding structure are described in Figure~\ref{fig:StegNetArch}. In the embedding procedure, the~cover image and the hidden image are concatenated by channel while only the embedded image is shown to the network. Two parts of the network are both majorly made up of one lifting layer which lifts from figure channels to a uniform of 32 channels, six~\(3 \times 3\) basic building blocks raising features into high-dimensional latent space and one reducing layer which transforms features back to image space.

\begin{figure}[H]
  \centering
  \includegraphics[width=0.8\linewidth]{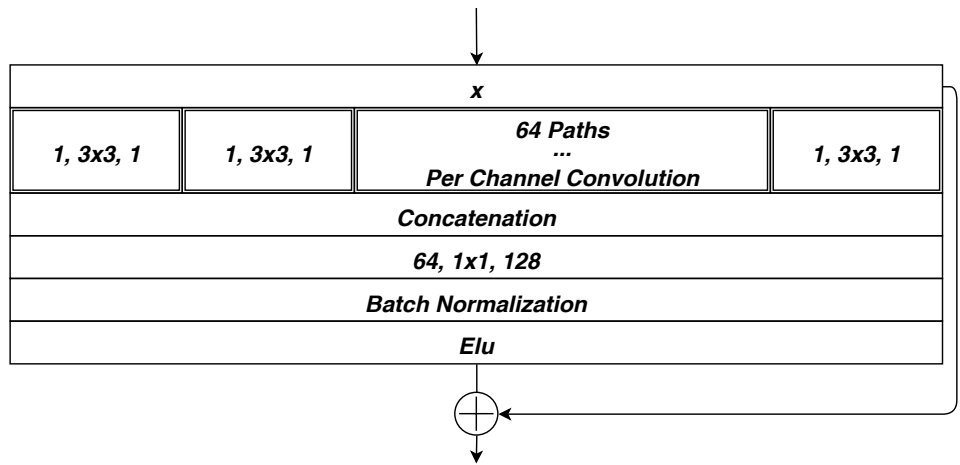}
  \vspace{-6pt}
  \caption{Separable Convolution with Residual Block.}%
\label{fig:SCRBlock}
\end{figure}

The basic building block named \textquote{Separable Convolution with Residual Block} (abbreviated as \textquote{SCR} in the following context) has the architecture as Figure~\ref{fig:SCRBlock}. We~adopt batch-normalization~\cite{batchnorm} and exponential linear unit (ELU)~\cite{elu} for quicker convergence and better result.

\subsection{Separable Convolution with Residual Block}%
\label{ssec:residualsepconv}

Our work adopt the state of the art neural network structure, the~skip connections in Highway Network~\cite{highway}, ResNet~\cite{resnet} and ResNeXt~\cite{resnext}, and separable convolution~\cite{sepconv} together to form the basic building block \textquote{SCR}.

\begin{figure}[H]
  \centering
  \includegraphics[width=0.7\linewidth]{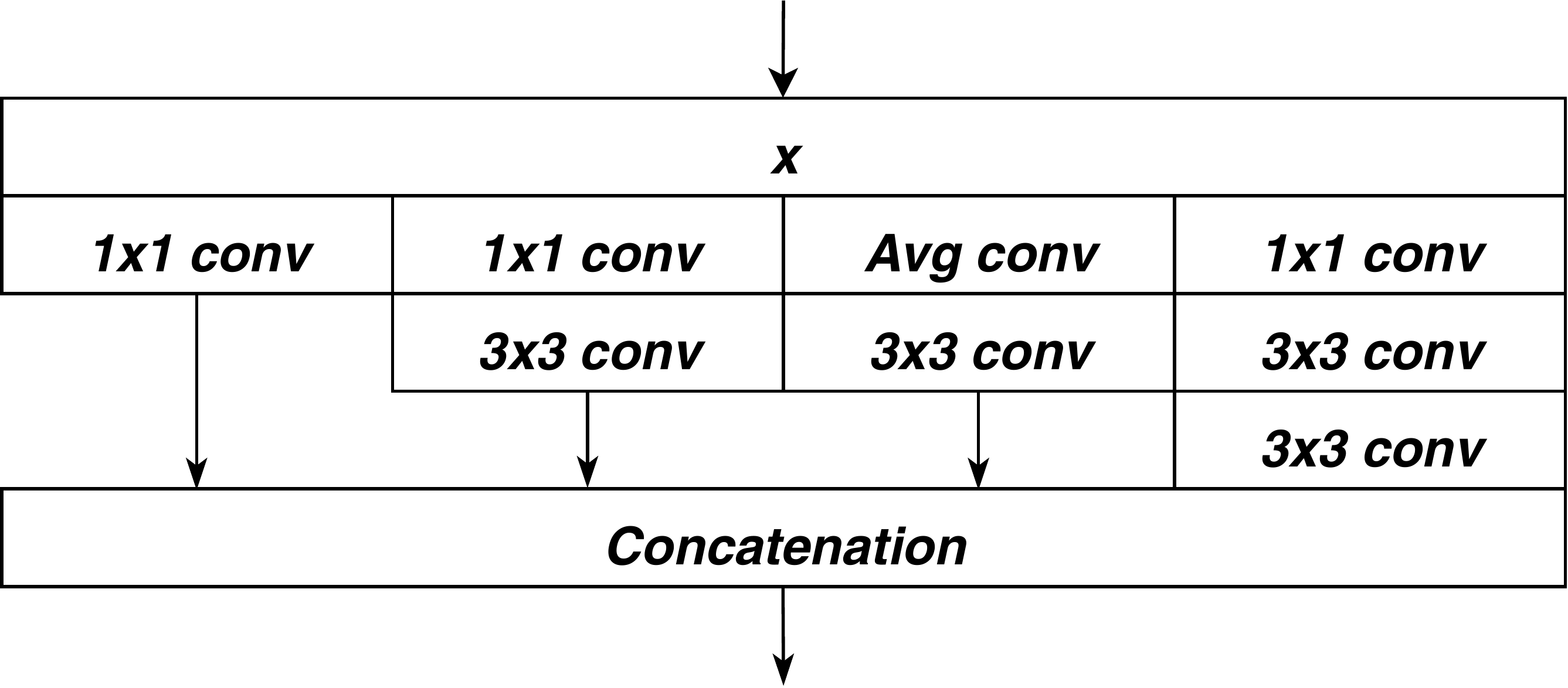}
\vspace{-3pt}
  \caption{Basic Building Block in Inception v3.}%
\label{fig:InceptionV3}
\end{figure}

The idea behind separable convolution~\cite{sepconv} originated from Google's Inception models~\cite{inceptionv1, inceptionv4} (see Figure~\ref{fig:InceptionV3} for its building blocks), and the hypothesis behind is that \textquote{cross-channel correlations and spatial correlations are decoupled}. Further, in Xception architecture~\cite{sepconv}, it makes an even stronger hypothesis that \textquote{cross-channel correlations and spatial correlations can be mapped completely separately}. Together with skip-connections~\cite{resnet} the gradients are preserved in backpropagation process via skip-connections to frontier layers and as a result, ease the problem of vanishing gradients.

\begin{figure}[H]
  \centering
  \begin{tabular}{cc}
    \adamIncludeFigure{0.3}{Embedding Structure}{structure/EncodeStructure}
    \adamIncludeFigure{0.3}{Decoding Structure}{structure/DecodeStructure}
  \end{tabular}
  \vspace{-11pt}
  \caption{StegNet Network Architecture.}%
\label{fig:StegNetArch}
\end{figure}
\vspace{-12pt}
  \vspace{-12pt}

\subsection{Training}%
\label{ssec:training}

Learning the end-to-end mapping function from cover and hidden image to embedded image and embedded image to decoded image requires the estimation of millions of parameters in the neural network. It is achieved via minimizing the weighted loss of \(L_{1}\)-loss between the cover and the embedded image, \(L_{1}\)-loss between the hidden and the decoded image, and their corresponding variance losses (variance should be computed across images' height, width and channel). \(C, H, E, D\) symbols stand for the cover image (\(C\)), the~hidden image (\(H\)), the~embedded image (\(E\)) and the decoded image (\(D\)) in correspondence. (See Equations~(\ref{eq:seploss1})--(\ref{eq:loss}))
\csdef{CE}{\mathrm{CE}}
\csdef{ED}{\mathrm{ED}}
\csdef{HD}{\mathrm{HD}}
\csdef{Var}{\mathrm{Var}}
\begin{align}
E_{i} &= F_{\CE}(C_{i}, H_{i}; \Theta_{\CE}) & D_{i} &= F_{\ED}(E_{i}; \Theta_{\ED})
\label{eq:seploss1}\\
L_{\CE} &= \frac{1}{n} \sum_{i=1}^{n} \abs{E_{i} - C_{i}} & L_{\HD} &= \frac{1}{n} \sum_{i=1}^{n} \abs{D_{i} - H_{i}}
\label{eq:seploss2}
\end{align}
\begin{equation} \label{eq:loss}
\mathrm{Loss} = \frac{1}{4} (L_{\CE} + L_{\HD} + \Var(L_{\CE}) + \Var(L_{\HD}))
\end{equation}

\(L_{\CE}\) is used to minimize the difference between the embedded image and the cover image, while~\(L_{\HD}\) is for the hidden image and the decoded image. Choosing only to decode the hidden image while not both the cover and the hidden images are under the consideration that the embedded image should be a concentration of high-level features apparently similar to the cover image whose dimension is half the shape of those two images, and some trivial information has been lost. It would have pushed the neural network to balance the capacity in embedding the cover and the hidden if both images are extracted at the decoding process.

Furthermore, adopting variance losses helps to give a hint to the neural network that the loss should be distributed throughout the image, but not putting at some specific position (See Figure~\ref{fig:VarianceDiff} for differences between. The~embedded image without variance loss shows some obvious noise spikes (blue points) in the background and also some around the dog nose).

\csundef{Var}
\csundef{ED}
\csundef{CE}

\begin{figure}[H]
  \centering
  \begin{tabular}{cc}
    \adamIncludeFigureCS{0.25}{0.45}{Embedded Image with Variance Loss}   {var_loss_effect/stegnet_00_04_steg_bbox}
    \adamIncludeFigureCS{0.25}{0.45}{Embedded Image without Variance Loss}{var_loss_effect/novar_00_04_steg_bbox} \\

    \adamIncludeFigureCS{0.25}{0.45}{Red Box Magnified (with Variance Loss)}   {var_loss_effect/stegnet_00_04_steg_mag}
    \adamIncludeFigureCS{0.25}{0.45}{Red Box Magnified (without Variance Loss)}{var_loss_effect/novar_00_04_steg_mag}
  \end{tabular}
  \vspace{-8pt}
  \caption{Variance Loss Effect on Embedding Results.}%
\label{fig:VarianceDiff}
\end{figure}

\section{Experiments}%
\label{sec:experiments}
\vspace{-6pt}
\subsection{Environment}%
\label{ssec:environment}

\begin{figure}[H]
  \centering
  \includegraphics[width=\linewidth]{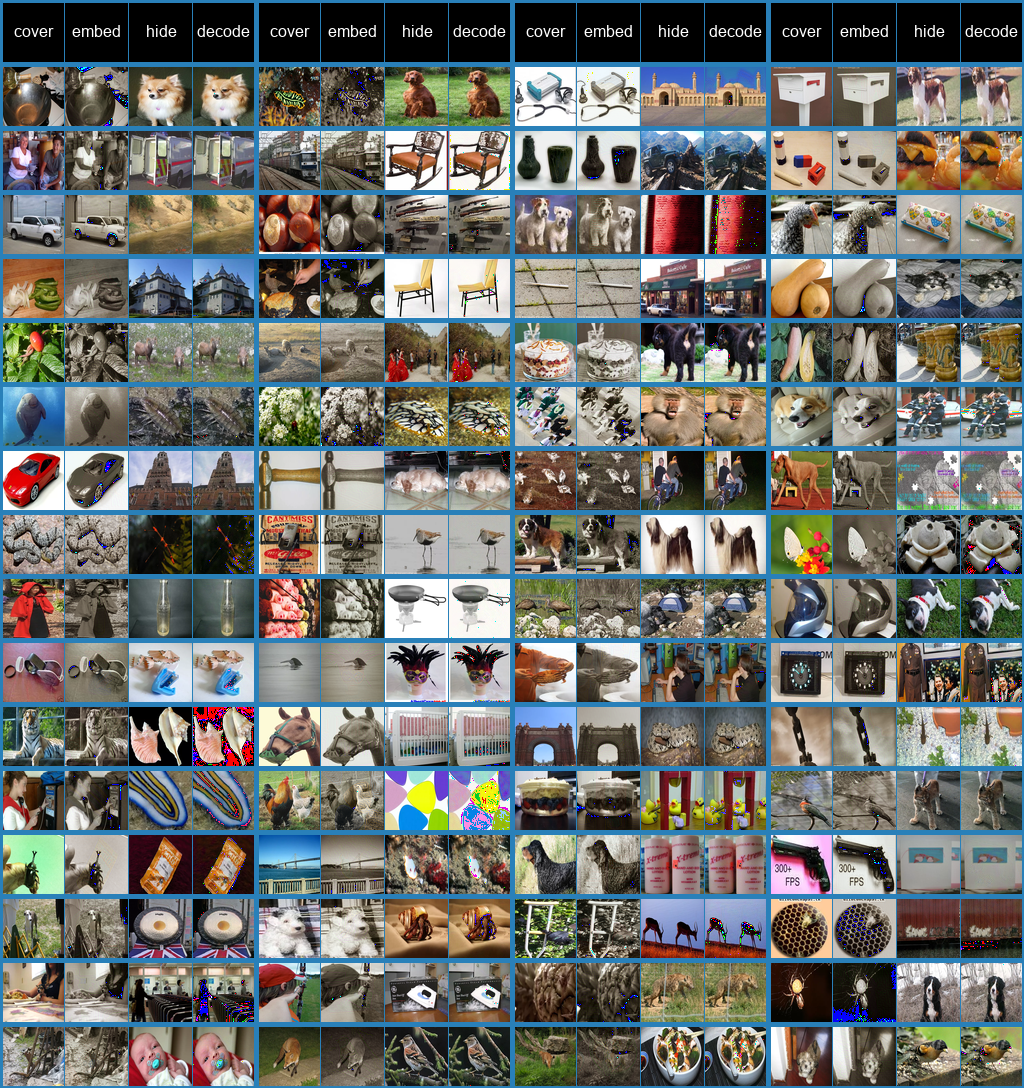}
  \caption{One batch Steganography Example.}%
\label{fig:onebatch}
\end{figure}

Our work is trained on one NVidia GTX1080 GPU and we adopt a batch size of 64 using Adam optimizer~\cite{adam} with learning rate at \(10^{-5}\). We~use no image augmentation and restrict model's input image to \(64 \times 64\) in height and width because of memory limit. Training with resized \(64 \times 64\) ImageNet can yield pretty good results. We~use 80\% of the ImageNet dataset for training and the remaining for testing to verify the generalization ability of our model. Figure~\ref{fig:onebatch} shows the result of applying StegNet steganography method on a batch of images. 

\subsection{Statistical Analysis}%
\label{ssec:statanalysis}

The encoded and decoded images comparison between StegNet and LSB method are presented in Figure~\ref{fig:stegnetvslsb3}. They are very similar though, however, there is one critical flaw about the LSB method, in that it does not suffer through statistical analysis, and therefore LSB method is usually combined with transformations of the hidden image, i.e.,\ compression, randomization, etc.

Figure~\ref{fig:stegnetvslsb3hist} is a comparison of histogram analysis between LSB method and our work. It shows a direct view of robustness of StegNet against statistical analysis, which~the StegNet embedded's histogram and the cover image's histogram are much more matching.

A more all-around test is conducted through StegExpose~\cite{stegexpose}, which~combines several decent algorithms to detect LSB-based steganography, i.e.,\ sample pair analysis~\cite{samplepairanalysis}, RS analysis~\cite{fridrich2004reliable}, chi-square attack~\cite{westfeld1999attacks} and primary sets~\cite{dumitrescu2002steganalysis}. The~detection threshold is its hyperparameter, which~is used to balance true positive rate and false positive rate of the StegExpose's result. The~test is performed with linear interpolation of detection threshold from 0.00 to 1.00 with 0.01 as the step interval.

\begin{figure}[H]
  \centering
  \includegraphics[width=0.82\linewidth]{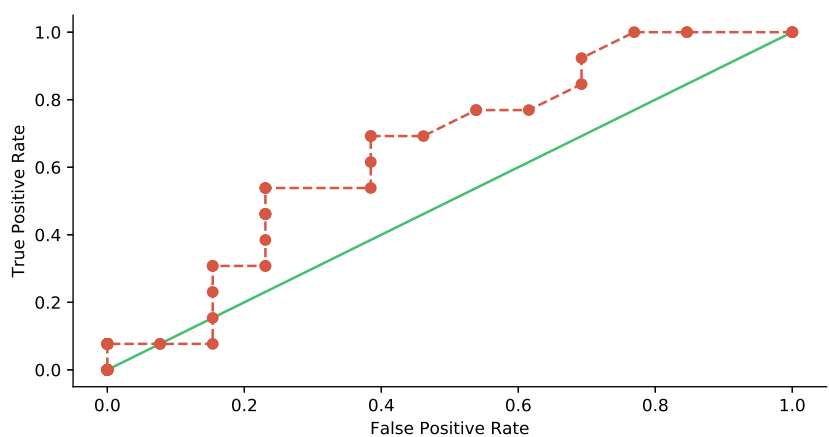}
  \vspace{-4pt}
  \caption{ROC Curves: Detecting Steganography via StegExpose.}%
\label{fig:roccurve}
\end{figure}

Figure~\ref{fig:roccurve} is the ROC curve, where true positive stands for an embedded image correctly identified that there are hidden data inside while false positive means a clean figure falsely classified as an embedded image. The~figure is plotted in red-dash-line-connected scatter data, showing that StegExpose can only work a little better than random guessing, the~line in green. In other words, the~proposed steganography method can better resist StegExpose attack.

\begin{figure}[H]
  \centering
  \begin{tabular}{ccc}
    \adamIncludeFigureCS{0.2}{0.3}{Cover Image}       {hist_analysis/image_1_covr.png}
    \adamIncludeFigureCS{0.2}{0.3}{StegNet Embedded}  {hist_analysis/image_1_steg_stegnet.png}
    \adamIncludeFigureCS{0.2}{0.3}{3-bit LSB Embedded}{hist_analysis/image_1_steg_lsb3.png} \\
  \vspace{-5pt}

    \adamIncludeFigureCS{0.2}{0.3}{Cover Image Histogram}{hist_analysis/image_1_covr_hist.png}
    \adamIncludeFigureCS{0.2}{0.3}{StegNet Histogram}    {hist_analysis/image_1_steg_stegnet_hist.png}
    \adamIncludeFigureCS{0.2}{0.3}{3-bit LSB Histogram}  {hist_analysis/image_1_steg_lsb3_hist.png}
  \end{tabular}
  \vspace{-8pt}
  \caption{Histogram Comparison between StegNet and Plain LSB.}%
\label{fig:stegnetvslsb3hist}
\end{figure}
\vspace{-12pt}

\section{Conclusions and Future Work}%
\label{sec:conclusion}

We have presented a novel deep learning approach for image steganography. We~show that the conventional image steganography methods mostly do not serve with good payload capacity. The~proposed approach, StegNet, creates an end-to-end mapping from the cover image, hidden~image to embedded image and from embedded image to decoded image. It has achieved superior performance than traditional methods and yet remains quite robust.

As seen in Figure~\ref{fig:onebatch}, there is still some noise generated at non-texture-rich areas, i.e.,\ plain white or plain black parts. The~variance loss adopted by StegNet might not be the optimal solution to loss~distribution.

In addition to the idea of ``trading accuracy for capacity'', the~embedded image does not need to be even visually similar to the cover image. The~only requirement to the embedded image is to pass the third-party supervision and the hidden image should be successfully decoded after the transmission is complete, and therefore the embedded image can look similar to anything that is inside the cover image dataset while can look nothing related to anything that is inside the hidden image dataset. Some of the state of the art generative models in neural networks can help achieve it, i.e.,\ Variational Autoencoders~\cite{VAE, AAE}, Generative Adversarial Networks~\cite{GAN, WGAN, BEGAN}, etc.

Some work is needed for non-equal sized images steganography since ``1:1'' image steganography is huge for traditional judgment; however, the~ability of neural networks still remains to be discovered. Whether it is possible to generate approaches for even better capacity, or with a better visual quality for even safer from detections. Some other work is needed for non-image hidden information steganography, i.e.,\ text information, binary data. In addition to changing the hidden information type, the~cover information type may also vary from text information to even videos. Furthermore, some~work is needed for applying StegNet on lossy-compressed image file formats or third-party spatial translations, i.e.,\ cropping, resizing, stretching, etc.

\vspace{6pt}

\authorcontributions{%
  Conceptualization, Y.Y. and X.L.; Data Curation: Y.Y.; Formal Analysis, Y.Y. and X.L.; Funding Acquisition, P.W. and X.L.; Investigation, Y.Y., X.L. and P.W.; Methodology, Y.Y. and X.L.; Project~Administration, P.W. and X.L.; Resources, Y.Y.; Software, Y.Y.; Supervision, P.W.; Validation, Y.Y.; Visualization, Y.Y.; Writing—Original Draft Preparation, Y.Y.; Writing—Review \& Editing, X.L. and Y.Y.}

\funding{%
This work was supported by the Shanghai Innovation Action Plan Project under grant number 16511101200.
}

\conflictsofinterest{%
The authors declare no conflict of interest.
}

\reftitle{References}

\end{document}